\documentstyle[prb,aps,twocolumn,epsf]{revtex}
\begin{document}
%------------------------------------------------------------------------------
%for pacs numbers print
\draft
%------------------------------------------------------------------------------
%?????????
\wideabs{
%?????????
\title{
 Crossover behavior of the $J_1$-$J_2$ model in a staggered magnetic
 field
}
\author{
 Hiromi Otsuka\cite{email}
}
\address{
 Department of Physics, Tokyo Metropolitan University, Tokyo 192-0397, Japan
}
\date{
 Received \today
}
%------------------------------------------------------------------------------
\maketitle
\begin{abstract}
 The ground states of the $S=\frac12$ Heisenberg chain with the
 nearest-neighbor and the next-nearest-neighbor antiferromagnetic
 couplings are numerically investigated in a staggered magnetic
 field.
 While the staggered magnetic field may induce the N\'eel-type
 excitation gap, and it is characterized by the Gaussian fixed point in
 the spin-fluid region, the crossover to the behavior controlled by the
 Ising fixed point is expected to be observed for the spontaneously
 dimerized state at finite field.
 Treating a low-lying excitation gap by the phenomenological
 renormalization-group method, we numerically determine the massless
 flow connecting the Gaussian and Ising fixed points. 
 Further, to check the criticalities, we perform the finite-size-scaling
 analysis of the excitation gap.
\end{abstract}
% 75.10.Jm   Quantized spin models 
% 64.60.Fr   Equilibrium properties near critical points, critical exponents  
%????????????
}\narrowtext
%????????????

%------------------------------------------------------------------------------
% INTRODUCTION
%------------------------------------------------------------------------------

 Theoretically and experimentally, there have been intensive
 investigations on critical phenomena observed in low-dimensional
 quantum spin systems.
 In particular, recent investigations on this subject have focused not
 only on critical fixed points and their neighboring regions, but also
 on the global behaviors of the renormalization-group (RG) flows
 connecting them.
 For one-dimensional (1D) quantum systems (also for 2D classical
 systems), besides the exact solutions available for some cases, the
 conformal field theory (CFT) provides the most efficient way to
 characterize the fixed points, where the values of the central
 charge $c$ specify their universality classes. 
 Further, with respect to the RG flows observed in the continuum and
 unitary models, Zamolodchikov's $c$-theorem serves for the explanations
 of their general properties.\cite{Zamo86}
 Since in investigations of the quantum spin chains and interacting
 electron systems, the $c=1$ CFT has close relevance to their
 criticality,\cite{Hald81} it is very important to understand its
 instability and a crossover to behaviors controlled by other critical
 fixed points. 

 In this paper, we shall study the ground states of the $S=\frac12$
 Heisenberg chain with the nearest-neighbor and the
 next-nearest-neighbor antiferromagnetic couplings (the so-called
 $J_1$-$J_2$ model) in a staggered magnetic field; the Hamiltonian to
 be considered is given by
 $H=H_1+H_2$ with
 \begin{eqnarray}
  &&H_1=
   \sum_{j=1}^L
   \left(
    2 J_1{\bf S}_j\cdot{\bf S}_{j+1}
    + 2 J_2{\bf S}_j\cdot{\bf S}_{j+2}
  \right),
   \label{eq-J1J2H}\\
  &&H_2=
   \sum_{j=1}^L
   -h_{\rm s} (-1)^jS^z_{j},
   \label{eq-MAGNH}
 \end{eqnarray}
 where $S^\nu_j$ is the $\nu$th component of the spin operator on the
 $j$th site ${\bf S}_j$ and couplings $J_1$, $J_2>0$ (we take $J_1$ as
 the energy unit in the following). 
 We assume the periodic boundary condition ${\bf S}_{L+1}={\bf S}_1$ and 
 an even number of $L$. 

 In the zero-field case ($h_{\rm s}=0$), there are two special points
 where the exact ground states have been known: at $J_2=0$ where the
 Bethe ansatz solution is available, the system is in the spin-fluid
 phase described by the Gaussian fixed point ($c=1$ CFT); at $J_2=0.5$
 (the Majumdar-Ghosh point) the direct products of spin singlet pairs
 formed either on bonds $\langle 2k,2k+1\rangle$ or $\langle
 2k+1,2k+2\rangle$ become twofold degenerated ground states (i.e.,
 spontaneously dimerized states),\cite{Maju70,Broe80,Shas81} and a
 finite excitation gap exists there.\cite{Affl88}
 According to Haldane\cite{Hald82}, and Kuboki and Fukuyama,\cite{Kubo87}
 the gap formation caused by the frustration can be well described by
 the quantum sine-Gordon model obtained via the bosonization
 procedure;\cite{Gogo98} the effective Hamiltonian of Eq.\
 (\ref{eq-J1J2H}) is given as
  \begin{eqnarray}
   {\cal H}_1
    &&=\int{d}x
    \frac{v}{2\pi}
    \left[
     {       K} \left(\partial_x \theta\right)^2 + 
     {1\over K} \left(\partial_x   \phi\right)^2  
   \right]\nonumber\\
    &&+
    \int{d}x
    \frac{2g_\phi}{(2\pi\alpha)^2}
    {\cos\sqrt8\phi},
   \label{eq-EF-H1}
  \end{eqnarray}
 where the bosonic operator $\theta$ is the dual field of $\phi$
 satisfying the commutation relation
 $
 \partial_y\left[\phi(x),\theta(y)\right]={\rm i}\pi\delta(x-y).
 $
 $K$ and $v$ are the Gaussian coupling and the spin-wave velocity, and
 $g_\phi$ stands for the spin Umklapp scattering bare
 amplitude.\cite{ORIGIN}
 Although the property of this effective continuum model has been well
 understood and actually the $g_\phi$ term may become
 relevant,\cite{Kogu79} for the determination of the fluid-dimer
 transition point $J_2^*$, numerical treatments of Eq.\ (\ref{eq-J1J2H})
 were required. 
 By carefully investigating the effect of the marginal $g_\phi$ term on
 the critical fixed point through lower-energy excitation levels
 observed in finite-size systems, Okamoto and Nomura precisely
 determined $J_2^*\simeq0.2411$.\cite{Okam92}  
 While it is known that the excitation gap also exists in the case of
 $J_2>0.5$,\cite{Tone87} we, in the following discussion, restrict
 ourselves to the region $0\le J_2\le0.5$ for simplicity.

 In the case of nonzero field
 [this is approximately realized in the quasi-1D antiferromagnets with
 the alternating gyromagnetic tensors, e.g., Cu-benzoate\cite{Oshi97}
 and Yb$_4$As$_3$\cite{Kohg99}
 and the field may be also generated as the intrinsic one originated
 from the N\'eel ordered sublattice\cite{Masl98}]
 since the bosonized form of Eq.\ (\ref{eq-MAGNH}) is given as
  \begin{equation}
   {\cal H}_2
    =\int{d}x
    -\frac{h_{\rm s}}{\pi\alpha}
    {\cos\sqrt2\phi},
   \label{eq-EF-H2}
  \end{equation}
 and the scaling dimension of this perturbation term is $x_2=\frac12$
 ($1/\nu=2-x_2=3/2$) on the fixed point,\cite{Kada79} the second-order
 phase transition occurs for systems in the spin-fluid region
 $J_2\le J_2^*$ with the divergence of correlation length $\xi\propto
 h_{\rm s}^{-2/3}$ (aside from the logarithmic correction due to the
 $g_\phi$ term).
 On the other hand, for $J_2>J^*_2$, due to the relevant $g_\phi$ term,
 the crossover of the transition driven by the staggered magnetic field
 occurs.
 Since the dimer gap may survive in a weak field, the critical point
 $h^*_{\rm s}(J_2)$ takes nonzero values depending on $J_2$, and further
 the universality class is changed.
 Recently, Fabrizio {\it et al.},\cite{Fabr00} on the basis of the
 double-frequency sine-Gordon (DSG) theory given by Delfino and
 Mussardo,\cite{Delf98} argued that the system on $h^*_{\rm s}(J_2)$ is
 renormalized to the Ising fixed point with $c=\frac12$ in accord with
 the ``downhill'' condition of the $c$-theorem, so the criticality in
 the vicinity of this line is related to the divergent correlation
 length of the form $\xi\propto [h_{\rm s}-h^*_{\rm s}(J_2)]^{-1}$
 (explained below) described by the $\varphi^4$ theory.

%------------------------------------------------------------------------------
% METHODOLOGY
%------------------------------------------------------------------------------

 From the viewpoint that the Gaussian fixed point is perturbed by two
 relevant operators of $g_\phi$ and $h_{\rm s}$, we can qualitatively
 estimate $h^*_{\rm s}(J_2)$ around the point according to the 
 crossover argument.\cite{Cardy}
%==========================
 However, to evaluate its precise value especially near the Ising fixed
 point, a numerical treatment of $H$ should be required.
 Here, it should be noted that the criticality on the N\'eel-phase
 boundaries in the $S=1$ bond-alternating {\it XXZ} chain also belongs
 to the Ising universality and they were precisely determined by the
 numerical method.\cite{Kita97}
 Thus, we shall perform our calculations by analyzing the lower-energy
 excitations observed in the finite-size systems. 
%==========================
 Let us first focus our attention to the excitations in the spin-fluid
 region described by the Gaussian fixed point,\cite{Nomu94,Nomu95} 
%
%==========================
 \begin{eqnarray}
  {\cal O}_{1}&=&\sqrt2\sin\sqrt2\phi,
   \label{eq_SPN1}     \\ 
  {\cal O}_{2}&=&\sqrt2\cos\sqrt2\phi,
   \label{eq_SPN2}     \\ 
  {\cal O}_{3}&=&{\rm exp(}\pm i\sqrt2\theta).
   \label{eq_SPN3}
 \end{eqnarray}
 ${\cal O}_{1}$ and ${\cal O}_{2}$ denote dimer and N\'eel excitations,
 respectively, while ${\cal O}_{3}$ is the doublet excitation changing
 an amount of the total spin.
 According to the finite-size-scaling (FSS) argument based on CFT,
 corresponding energy levels $\Delta E_i$  for these operators (taking
 the ground-state energy as zero) are expressed by the use of their
 scaling dimensions $x_{i}$ as $\Delta E_{i}\simeq 2\pi v
 x_{i}/L$.\cite{Card84}
 For the case of $h_{\rm s}=0$, we can calculate $\Delta E_i$ according
 to the level-spectroscopy method,\cite{Nomu95} where discrete
 symmetries of the lattice Hamiltonian (see below) are utilized to
 specify excitation levels.  
 On the other hand, for the case of $h_{\rm s}\ne0$, the usable symmetry
 becomes lower, and more importantly, the universality class is changed
 to the Ising one on the line $h^*_{\rm s}(J_2)$ so that we should
 employ other criterion to characterize the levels.
 Here, we will use the so-called UV-IR (ultraviolet-infrared) operator
 correspondence,\cite{Fabr00,Bajn01} i.e., along the critical RG flow
 the operators on the Gaussian fixed point (UV) are transmuted to those
 on the Ising fixed point (IR) as

%FFFFFFFFFFFFFFFFFFFFFFFFFFFFFFFFFFFFFFFFFFFFFFFFFFFFFFFFFFFFFFFFFFFFFFFFFFFFFF
 \begin{figure}[t]
%???????????????
% \psbox[vscale=0.5,hscale=0.5]{fig1.eps}
  \centerline{\epsfysize=60mm \epsffile{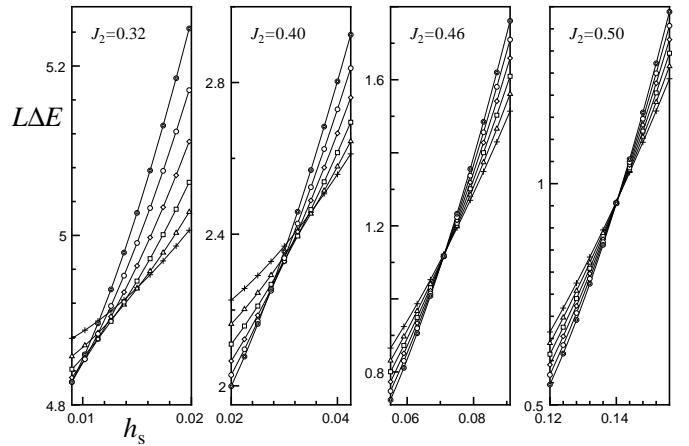}}
%???????????????
  \caption{
 The $L$ and $h_{\rm s}$ dependences of $L\Delta E(J_2,h_{\rm s},L)$. 
 From left to right, the next-nearest-neighbor coupling $J_2$=0.32, 0.40,
  0.46, and 0.5. 
 Data for systems of $L=18$ (crosses), 20 (triangles), 22 (squares), 24
  (diamonds), 26 (circles), and 28 (double circles) are plotted with the
  fitting curves.
  }
  \label{FIG1}
 \end{figure}
%FFFFFFFFFFFFFFFFFFFFFFFFFFFFFFFFFFFFFFFFFFFFFFFFFFFFFFFFFFFFFFFFFFFFFFFFFFFFFF

%
 \begin{equation}
  {\cal O}_{1} \to \mu,~~~{\cal O}_{2} \to I+\epsilon,
   \label{eq-UV-IR}
 \end{equation}
 where $\mu$ is the disorder field (Z$_2$ odd) and $\epsilon$ is the
 energy density operator (Z$_2$ even) with scaling dimensions
 $x_\mu=\frac18$ and $x_\epsilon=1$, respectively.
 According to this correspondence, we can obtain a relevant excitation
 in nonzero field by taking the limit $h_{\rm s}\searrow 0$ and
 assigning it to one of well-characterized states.
 From Eqs. (\ref{eq-EF-H2}) and (\ref{eq-UV-IR}), the staggered magnetic
 field plays a role of the ``thermal'' scaling variable on $h^*_{\rm s}(J_2)$,
 and thus the critical exponent $1/\nu=2-x_\epsilon=1$ as already
 mentioned (see Ref.\ \onlinecite{Fabr00}).  
 On one hand, since the ``magnetic'' excitation $\mu$ stemming from the
 dimer excitation provides lower energy (i.e., the most divergent
 fluctuation), we will thus focus attention on it for the determination
 of $h^*_{\rm s}(J_2)$.

 Now, we shall explain the numerical calculation procedure.
 Since the system is massive unless it is located on the critical line,
 the phenomenological renormalization-group (PRG) method is expected to
 work efficiently for our aim:\cite{Room80}
 We numerically solve the PRG equation for the systems of $L$ and $L+2$, 
 \begin{equation}
  (L+2)\Delta E(J_2,h_{\rm s},L+2)=L\Delta E(J_2,h_{\rm s},L)
   \label{eq-PRG}
 \end{equation}
 with respect to $h_{\rm s}$ for a given value of $J_2$. 
 Since the equation can be satisfied by the gap having the size
 dependence of $\Delta E(J_2,h_{\rm s},L)\propto 1/L$, the obtained
 value can be regarded as the $L$-dependent transition point, say
 $h^*_{\rm s}(J_2,L+1)$. 
 Then, extrapolating them to the thermodynamic limit, we estimate
 $h^*_{\rm s}(J_2)$.
 Alternatively, as explained in Refs.\ \onlinecite{Nomu94} and
 \onlinecite{Nomu95}
 ---
 to which we refer the interested readers for a detailed explanation
 ---
 at $h_{\rm s}=0$,
 the dimer excitation is in the subspace of
 total spin $S^z_{\rm T}=0$,
 wave number $k=\pi$,
 space inversion $P=+1$, and
 spin reversal $T=+1$
 for $L\equiv0$ (mod 4)
 [$S^z_{\rm T}=0$,
 $k=0$,
 $P=-1$, and
 $T=-1$
 for $L\equiv2$ (mod 4)]. 
 Thus, we can calculate
 $\Delta E(J_2,h_{\rm s},L)=E_\mu(J_2,h_{\rm s},L)-E_{\rm g}(J_2,h_{\rm s},L)$
 from
 $E_\mu(J_2,h_{\rm s},L)$
 connecting to the dimer excitation and
 $E_{\rm g}(J_2,h_{\rm s},L)$, i.e., 
 the ground-state energy. 

%------------------------------------------------------------------------------
% NUMERICAL RESULTS
%------------------------------------------------------------------------------

%FFFFFFFFFFFFFFFFFFFFFFFFFFFFFFFFFFFFFFFFFFFFFFFFFFFFFFFFFFFFFFFFFFFFFFFFFFFFFF
 \begin{figure}[t]
%???????????????
%  \begin{center}
%   \psbox[vscale=0.52,hscale=0.52]{fig2.eps}
%  \end{center}
  \centerline{\epsfysize=60mm \epsffile{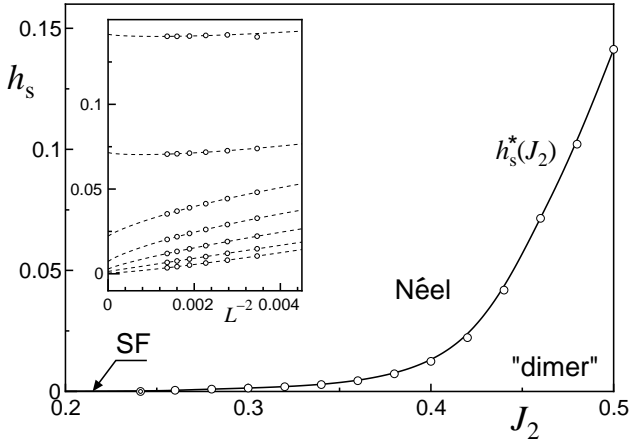}}
%???????????????
  \caption{
  The boundary line of the N\'eel and ``dimer'' phases (for an
  explanation of ``dimer'' phase, see
  Ref.\ \protect{\onlinecite{Fabr00}}). 
  Inset shows some of extrapolations of the $L$-dependent critical
  fields to the thermodynamic limit, where fitting curves are given.
  The double circle shows the fluid-dimer transition
  point $(J_2,h_{\rm s})=(J^*_2,0)$ at which the criticality of the
  boundary changes from the Gaussian to the Ising type.
  The spin-fluid (SF) state exists on the $x$ axis of $J_2\le J^*_2$. 
  }
  \label{FIG2}
 \end{figure}
%FFFFFFFFFFFFFFFFFFFFFFFFFFFFFFFFFFFFFFFFFFFFFFFFFFFFFFFFFFFFFFFFFFFFFFFFFFFFFF
%FFFFFFFFFFFFFFFFFFFFFFFFFFFFFFFFFFFFFFFFFFFFFFFFFFFFFFFFFFFFFFFFFFFFFFFFFFFFFF
 \begin{figure}[t]
%???????????????
%  \begin{center}
%   \psbox[vscale=0.52,hscale=0.52]{fig3.eps}
%  \end{center}
  \centerline{\epsfysize=85mm \epsffile{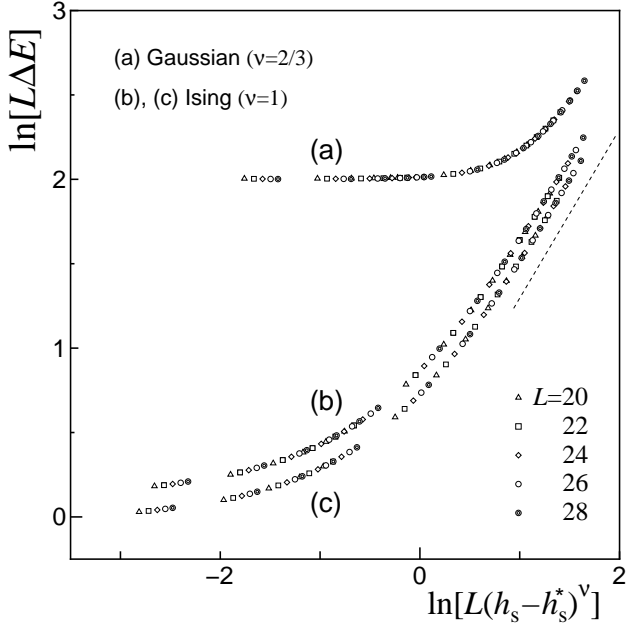}}
%???????????????
  \caption{
  The FSS plots of the excitation gap for systems of $L=20-28$ in
  log-log scale.
  We have used $\nu=2/3$ for the Gaussian transition [(a) $J_2=J^*_2$], and 
 $\nu=1$ for the Ising transition [(b) $J_2=$0.46, (c) $J_2=$0.5]. 
  The slope of the dotted line is 1 showing the expected asymptotic
  behavior of the scaling functions.  
  }
  \label{FIG3}
 \end{figure}
%FFFFFFFFFFFFFFFFFFFFFFFFFFFFFFFFFFFFFFFFFFFFFFFFFFFFFFFFFFFFFFFFFFFFFFFFFFFFFF

 In Fig.\ \ref{FIG1}, we demonstrate $L$ and $h_{\rm s}$ dependences of
 $L\Delta E(J_2,h_{\rm s},L)$ at several values of $J_2$.
 Systems up to $L=28$ sites are treated, where the Lanczos algorithm is
 used to obtain eigenvalues of the Hamiltonian in specified subspaces.
 We can see that the $L$ dependence of the crossing point is almost
 absent for large value of $J_2$ (near $J_2=0.5$), while it is visible
 for the small value case.

%==========================
 After evaluating $h^*_{\rm s}(J_2,L+1)$, we extrapolate them to
 $L\to\infty$; here we assume the following formula: 
 \begin{equation}
  h^*_{\rm s}(J_2,L)=h^*_{\rm s}(J_2)+a L^{-2} (1+b L^{-\omega})~~~
   (\omega>0),
   \label{eq-EXTRAP}
 \end{equation}
%==========================
 where the $b$ term stands for a correction to the leading one, 
 and four parameters $h^*_{\rm s}(J_2)$, $a$, $b$, and $\omega$ are
 determined according to the least-square-fitting condition.
 We used the data of $L=18-28$, and extrapolated them to $h^*_{\rm
 s}(J_2)$ as shown in the inset of Fig.\ \ref{FIG2}, where from bottom
 to top a series of data with fitting curves are given in the increasing
 order of $J_2$.
 Consequently, we obtain the critical line $h^*_{\rm s}(J_2)$ as shown
 in the figure.
 The RG eigenvalue of the scaling variable $g_\phi$ is ``almost
 zero''(i.e., marginal), while that of the staggered magnetic field is
 3/2 on the point $(J_2,h_{\rm s})=(J^*_2,0)$.
 The phase boundary is thus expected to behave as 
 $h^*_{\rm s}(J_2)\propto(J_2-J^*_2)^{1.5/0_+}$, which agrees with a weak $J_2$
 dependence of the line near the point.

 At this stage, we shall check the criticalities by the use of the
 FSS analysis in the vicinity of the boundary lines,\cite{Barb83} i.e.,
 we assume the following one-parameter scaling form of the excitation
 gaps in the finite system of $L$:
 \begin{equation}
  \Delta E(J_2,h_{\rm s},L)=L^{-1}\Psi(L(h_{\rm s}-h^*_{\rm s}(J_2))^\nu), 
   \label{eq-FSS}
 \end{equation}
 where
 $\nu=2/3$ for $J_2\le J^*_2$ and $\nu=1$ for $J_2>J^*_2$. 
 Further, the asymptotic behaviors of the scaling function are expected
 as
 $\Psi(x)\propto x$ for large $x$ and 
 $\Psi(x)\simeq {\rm const}$ for $x\to0$. 
 Using the obtained transition points, we plot Eq.\ (\ref{eq-FSS}) for
 the Gaussian $(J_2=J_2^*)$ and the Ising transitions ($J_2=$0.46 and
 0.5) in Fig.\ \ref{FIG3}.
 The results show that the data of different system sizes are collapsed
 on the single curve in both cases, and the asymptotic behaviors of
 $\Psi$ agree with the expected ones (dotted lines) despite of the small
 $L$.
 From these plots, we can check the crossover behavior of the
 transitions driven by the staggered magnetic field in the present
 frustrated quantum spin chain system and also the accuracy of the phase
 boundary line in Fig.\ \ref{FIG2}.
 This good FSS behaviors may rely on the conditions that the marginal
 operator that brings about the multiplicative logarithmic correction
 to the pure power-law singularity of $\xi$ is absent on the fluid-dimer
 transition point,\cite{Otsu98} and that the crossover region for
 $J_2>0.4$ may be large enough to be detected.
 On the other hand, the FSS nature becomes obscure in the weak and
 intermediate values of $J_2$ since our system sizes are too small to
 reach the region. 

 Finally, we evaluate the values of the central charge through the $L$
 dependence of the ground-state energy:\cite{Blot86} 
 \begin{equation}
  E_{\rm g}(J_2,h^*_{\rm s}(J_2),L)\simeq e_0 L - \frac{\pi vc}{6L}
   \label{eq-CENTRAL}
 \end{equation}
 ($e_0$ is the energy density in the thermodynamic limit).
 For this calculation, we should estimate $v$ in advance; here we use
 the FSS relation $\Delta E_\mu\simeq 2\pi vx_\mu(=\frac18)/L$ and the
 least-square-fitting procedure for the data.
 The obtained results agree well with the Ising one for the considerably
 large values of $J_2$, i.e.,
 $(v,c)\simeq$
 (1.439,0.494) at $J_2$=0.46 and
 (1.248,0.491) at $J_2$=0.50. 
 For comparison, we also estimate them at $J_2=J^*_2$ and obtain 
 $(v,c)\simeq$
 (2.365,0.995)
 [here the relation $\Delta E_1\simeq 2\pi vx_1(=\frac12)/L$ has been
 used to estimate $v$]. 
 Consequently, we can again confirm the universalities of critical
 systems on the phase boundary line.
 On the other hand, we could not extract the reliable data using the
 above procedure in the small and intermediate regions of $J_2$; this
 may be due to the finite-size effects on the line so that more detailed
 analysis might be required in this region.

%------------------------------------------------------------------------------
% SUMMARY
%------------------------------------------------------------------------------

 To summarize, we have numerically investigated the ground-state phase
 diagram of the one-dimensional $J_1$-$J_2$ model in the staggered
 magnetic field. 
 The crossover behavior of the second-order phase transitions driven by
 the staggered magnetic field occurs between the Gaussian and the
 Ising fixed points. 
 According to the operator correspondence between these fixed points and
 using the level-spectroscopy technique, we have analyzed the $Z_2$-odd
 excitation gap by the use of the phenomenological renormalization-group
 method, and determined the boundary line $h^*_{\rm s}(J_2)$ representing
 the massless flow connecting the fixed points.
 To check the two criticalities, we have performed the
 finite-size-scaling analysis of the excitation gap for typical
 parameter values, and we have also evaluated the values of the central
 charge.

 The present investigation has based upon the recent development of the
 (1+1)-dimensional double-frequency sine-Gordon theory, where its very
 interesting applications in wide areas of researches including
 condensed matter physics have been pointed out.\cite{Fabr00,Delf98} 
 However, as in the present case, numerical treatments of finite-size
 systems may be required for quantitative discussions on the lattice
 Hamiltonian models.
 We think that our numerical approach can also serve for the investigations
 of more complicated systems such as the correlated
 electrons;\cite{Fabr99,Tsuc99} we will report on the application results
 of this approach in the future publications.

%------------------------------------------------------------------------------
\section*{ACKNOWLEDGMENTS}
%------------------------------------------------------------------------------

 The author is grateful to
 M. Sumitomo
 and
 Y. Okabe
 for helpful discussions.
 Main computations were performed using the facilities of
 Tokyo Metropolitan University,
 Yukawa Institute for Theoretical Physics, 
 and 
 the Supercomputer Center, Institute for Solid State Physics, University
 of Tokyo.

%------------------------------------------------------------------------------

%------------------------------------------------------------------------------

\begin{references}
\bibitem{ORIGIN}
 The origin of the phase variable $\phi$ has been determined so
 as to $g_\phi<0$ in the $J_2=0$ case (cf. Ref.\onlinecite{Fabr00}).
\end{references}
\end{document}